\documentclass[pra,amsmath,aps,twocolumn]{revtex4}

\usepackage{graphicx}
\usepackage{rotating}

\newcommand{\bal}{\begin{align}}
\newcommand{\eal}{\end{align}}

\DeclareMathOperator{\trace}{Tr}
\newcommand{\comm}[2]{\ensuremath{\left[#1, #2\right]}}

\newcommand{\vv}[1]{\tilde{#1}}
\newcommand{\Hint}{H_{\text{int}}}
\newcommand{\Hintt}{\tilde{H}_{\text{int}}}
\newcommand{\rhot}{\tilde{\rho}_\textrm{t}}

\newcommand{\be}{\begin{equation}} \newcommand{\ee}{\end{equation}}

\newcommand{\ie}{i.e.}
\newcommand{\eg}{e.g.}
\newcommand{\rhodq}{\rho}
\newcommand{\envop}{E}
\begin{document}

\title{Equivalent qubit dynamics under classical and quantum noise}
\author{O.-P.~Saira,$^{1,2,}$\footnote[1]{Electronic address: \texttt{ops@fyslab.hut.fi}} V.~Bergholm,$^{1}$ T.~Ojanen,$^{2}$ and M.~M\"ott\"onen$^{1,2}$}
\affiliation{$^{1}$Laboratory of Physics, Helsinki University of Technology \\ P.~O.~Box 4100, FI-02015 TKK, Finland \\
$^{2}$Low Temperature Laboratory, Helsinki University of Technology \\ P.~O.~Box 2200, FI-02015 TKK, Finland}

\date{\today}

\begin{abstract}
We study the dynamics of quantum systems under classical and quantum noise, focusing on decoherence in qubit systems. Classical noise is
described by a random process leading to a stochastic temporal evolution of a closed quantum system, whereas quantum noise originates from the
coupling of the microscopic quantum system to its macroscopic environment. We derive deterministic master equations describing the average
evolution of the quantum system under classical continuous-time Markovian noise and two sets of master equations under quantum noise.
Strikingly, these three equations of motion are shown to be equivalent in the case of classical random telegraph noise and proper quantum
environments. Hence fully quantum-mechanical models within the Born approximation can be mapped to a quantum system under classical noise.
Furthermore, we apply the derived equations together with pulse optimization techniques to achieve high-fidelity one-qubit operations under
random telegraph noise, and hence fight decoherence in these systems of great practical interest.
\end{abstract}

\hspace{5mm}

\maketitle

\section{Introduction}
A quantum computer is an emerging computational device superior to its classical counterpart in certain tasks of great practical significance,
\eg, factorization of large integers~\cite{shor94}, unsorted database search~\cite{Grover} and the simulation of quantum
systems~\cite{Feynman,Alan2005}. The main concern in building a working large-scale quantum computer is the undesired coupling of quantum bits,
qubits, to their environment. This coupling results in errors and loss of purity, a phenomenon generally referred to as decoherence. However, an
important branch of quantum information science, fault tolerant quantum computing~\cite{shor1996}, states that if the qubit operations can be
achieved with an error probability below a certain constant threshold, the total computational error can actually be rendered arbitrarily small.
Thus studies on the effects of noise and decoherence on the qubit dynamics and especially direct suppression of the induced errors in qubit
operations are of great interest.

Recently, dynamical decoupling of quantum systems from their environments has been under active research. In the original
scenario~\cite{viola1998}, hard, infinitely fast control pulses were applied to the qubit to effectively turn off the coupling to its
environment, and hence to preserve its state by suppressing decoherence. It has been shown that it is also possible to generate qubit operations
without disturbing the decoupling process~\cite{viola1999} and that fast soft pulses suffice for the control~\cite{viola2003}. Furthermore, the
effectiveness of dynamical decoupling has been improved by introducing randomness to the control~\cite{viola2005a,viola2005b}. On the other
hand, a direct pulse optimization method to obtain high-fidelity qubit operations in the presence of classical random telegraph noise (RTN) in
the qubit energy splitting was reported in Ref.~\cite{mottonen2006}.

In this paper, we derive deterministic master equations for the average temporal evolution of a quantum system under classical continuous-time
Markovian noise. In the case of a qubit under RTN, our equations reduce to two coupled one-qubit master equations. This formulation allows for
great speed-up in dynamical decoupling as well as direct pulse optimization schemes for high-fidelity quantum gates compared with averaging over
many noise realizations as was done in Ref.~\cite{mottonen2006}. Since the master equations are not stochastic, they are also well suited for
algebraic manipulations.

For quantum noise, we study the dynamics of a system coupled to a macroscopic quantum environment within the standard Born approximation. In
addition, we present another set of master equations for a system under noise due to coupling to an impurity which is furthermore coupled to a
Markovian environment. The latter case has been extensively studied, for example, in Refs.~\cite{Paladino2002,Falci2002,Falci2005} for
suppression of random telegraph and $1/f$ noise in solid state qubits. Even though the obtained master equations for quantum noise are already
known, we demonstrate an interesting connection between the different approaches---the master equations for classical noise are equivalent to
those obtained in the case of quantum noise for certain couplings of the system to the environment. This result shows how classical noise can
arise from coupling to a quantum bath, and that in some cases classical noise can accurately model quantum noise. We stress that our master
equations yield the complete dynamics of the quantum system under noise in the Born approximation, not only for example damping rates studied in
Ref.~\cite{schoekopf}.

\section{Closed quantum system under classical Markovian noise}

We begin by considering a quantum system influenced by classical continuous-time Markovian noise with $N$ discrete states, each corresponding to
a different system Hamiltonian. The average outcome of any measurement applied to the system is obtained from~$\rho(t)$, the density operator
averaged over all possible noise trajectories. Although each noise trajectory alone corresponds to unitary evolution, the averaged evolution can
be nonunitary.

The Markovian assumption implies that the noise has no memory, \ie, at time~$t$ the state of the noise process is fully described by the
probabilities~$P_k(t)$ of the noise states, and the probability flow is governed by the transition rate matrix $\gamma=\{\gamma_{kj}\}$ as
\be\label{nprob}
\partial_t P_k(t) = \sum_{j=1}^N \gamma_{kj} P_j(t),
\ee where $\sum_k P_k(t) = 1$.
 We define the operator $\rho_k(t)$ as the density operator averaged over all noise
trajectories occupying the state~$k$ at time instant~$t$, and normalized such that $\trace[\rho_k(t)] = P_k(t)$. The total state operator of the
system is obtained as \be \label{densm} \rho(t) = \sum_{j=1}^N \rho_k(t). \ee The evolution of each $\rho_k(t)$ under an infinitesimal time
interval is given by the sum of all possible quantum evolutions weighted by their probabilities. Thus we obtain a closed set of ordinary master
equations for the conditional density operators~$\rho_k(t)$:
\begin{align}
\label{meq1}
\partial_t \rho_k(t) = \frac{1}{i \hbar} \comm{H_k(t)}{\rho_k(t)} +\sum_{j=1}^N \gamma_{kj} \rho_j(t),
\end{align}
where~$H_k(t)$ is the system Hamiltonian corresponding to the $k$th noise state. One can verify the consistency of Eqs.~\eqref{nprob}
and~\eqref{meq1} by taking a trace of Eq.~\eqref{meq1}, which results in Eq.~\eqref{nprob}. We point out that the conditional density
operators~$\rho_k(t)$ are introduced only for calculational purposes to obtain the average density operator~$\rho(t)$ according to
Eq.~\eqref{densm}. For RTN, \ie, a single bistable fluctuator, Eqs.~\eqref{meq1} reduce to
\begin{align} \partial_t \rho_\pm(t) =& \frac{1}{i \hbar}
\comm{H_\pm(t)}{\rho_\pm(t)}\pm\frac{1}{\tau_\textrm{c}}(\rho_--\rho_+), \label{rtn_master} \end{align} where the correlation time of the noise
is denoted by~$\tau_\textrm{c}$. Classical RTN similar to Eq.~\eqref{rtn_master} has been previously studied for example in the case of
macroscopic quantum tunneling in Josephson devices~\cite{Ankerhold2000}. We apply Eq.~\eqref{rtn_master} to finite-dimensional qubit systems in
Sec.~\ref{sec5}.

\section{Open quantum system in the Born approximation}
Let us turn our attention to quantum noise by studying a quantum system coupled to its environment. The temporal evolution of the total density
operator $\rho_\textrm{t}$ is governed by the Hamiltonian \be\label{hami} H = \underbrace{H_\textrm{s} \otimes I_\textrm{e} + I_\textrm{s}
\otimes H_\textrm{e}}_{=:H_0} + \underbrace{g K \otimes \envop}_{=:\Hint},\ee where the pure system and environmental Hamiltonians are denoted
by~$H_\textrm{s}$ and~$H_\textrm{e}$, respectively, and the interaction of the system with the environment is described by~$H_\textrm{int}$.
Without loss of generality, we can assume that the expectation value of the operator~$\envop$ over the environmental degrees of freedom
vanishes. To obtain the lowest-order corrections in the coupling strength~$g$ to the temporal evolution of the system, we write the master
equation for~$\rho_\textrm{t}$ in the interaction picture with respect to $H_0$ as $\partial_t \rhot = \frac{1}{i \hbar} [\Hintt,\rhot]$, where
$\rhot := U_0^\dagger(t) \rho_\textrm{t} U_0(t)$ and $\Hintt := U_0^\dagger(t) \Hint U_0(t)$. The evolution operator~$U_0(t)$ is expressed using
the time ordering operator~$\mathcal{T}$ as $U_0(t) = \mathcal{T} e^{\frac{1}{i\hbar} \int_0^t H_0(s) \mathrm{d}s}$. The master equation can be
expressed equivalently in an integro-differential form as~\cite{charmichael}
\begin{align}\label{master}
\partial_t \rhot &= \frac{1}{i \hbar} \comm{\Hintt(t)}{\rhot(0)}
\\ \nonumber &+\frac{1}{(i \hbar)^2} \int_0^t \comm{\Hintt(t)}{\comm{\Hintt(s)}{\rhot(s)}} \mathrm{d}s.
\end{align}
We proceed by assuming that the coupling of the system to the environment is sufficiently weak for the Born approximation to be valid, \ie,
$\rho_\textrm{t} \approx \rho_\textrm{s} \otimes \rho_\textrm{e}$. Thus the first term on the right hand side of Eq.~\eqref{master}, containing
operator~$\envop$ only to the first order, vanishes in taking the trace over the environmental degrees of freedom. The remaining part of the equation
yields
\begin{align}\label{qmaster1}
\partial_t \vv{\rho}_\textrm{s} &= \frac{g^2}{(i \hbar)^2} \int_0^t \left\{ \comm{\vv{K}(t)}{\vv{K}(s) \vv{\rho}_\textrm{s}(s)} C(t-s,0)\right. \nonumber\\
&\left.+\comm{\vv{K}(t)}{-\vv{\rho}_\textrm{s}(s) \vv{K}(s)}C(0,t-s) \right\} \mathrm{d}s,
\end{align}
where the forward and backward autocorrelation functions are defined as $C(t-s,0)=\trace\{\vv{\envop}(t-s) \vv{\envop}(0) \rho_\textrm{e}\}$ and
 $C(0,t-s)=\trace\{\vv{\envop}(0) \vv{\envop}(t-s) \rho_\textrm{e}\}$.

Since the operators~$\vv{\envop}$ at different instants of time do not necessarily commute, Eq.~\eqref{qmaster1} cannot in general be
simplified. However, for environments at high enough temperatures~$T$ the Kubo-Martin-Schwinger boundary condition
$C(t-s,0)=C[0,t-s+i\hbar/(k_\textrm{B}T)]$ yields
$C(t-s,0)\approx C(0,t-s)=:C_\textrm{e}(t-s)$, and hence Eq.~\eqref{qmaster1} reduces to
\begin{align}
\partial_t \vv{\rho}_\textrm{s} &=
\frac{g^2}{(i \hbar)^2} \int_0^t C_\textrm{e}(t-s) \comm{\vv{K}(t)}{\comm{\vv{K}(s)}{\vv{\rho}_\textrm{s}(s)}} \mathrm{d}s. \label{ID}
\end{align}
For example, quantum noise arising from a trapping center hybridized with the Fermi sea in a superconductor at temperatures much higher than the
linewidth of the trap results in RTN~\cite{desousa05} with the autocorrelation function $C_\textrm{e}(t-s) = e^{-2\frac{|s - t|}{\tau_c}}$.
Surprisingly, insertion of this correlation function into Eq.~(\ref{ID}) results in the same temporal evolution of the average density operator
as the master equations~\eqref{rtn_master} for classical RTN. This equality can be formally verified by showing that if Eq.~\eqref{ID} holds,
\be \label{yrite}\rho_\pm = \frac{1}{2} \vv{\rho}_\textrm{s} \pm \frac{g}{2 i \hbar} \int_0^t e^{-2\frac{t-s}{\tau_c}}
\comm{\vv{K}(s)}{\vv{\rho}_\textrm{s}(s)} \mathrm{d}s \ee is a solution to Eq.~(\ref{rtn_master}) with $H_\pm = \pm g \tilde{K}$, and that
$\rho_+ + \rho_- = \tilde{\rho}_\textrm{s}$. Thus we have shown that quantum noise arising from a realistic environment can be accurately
modeled with classical noise within the Born approximation. We note that although Eq.~\eqref{ID} seemingly includes memory effects, i.e., the
time derivative of the density matrix at time instant~$t$ depends on the density matrix at earlier times, the above proved equivalence shows
that no memory is required in the case of RTN if the dimension of the differential equation system is doubled.

The above approach can also be employed when the influence of the environment is modeled classically by including a stochastic noise term in the
system Hamiltonian. A stochastic system Hamiltonian analogous to the Hamiltonian in Eq.~(\ref{hami}) reads $H = H_\textrm{s} + g \xi(t) K$,
where~$\xi(t)$ is determined from some stochastic process. The resulting equations are similar to those presented above, except that each
expectation value over the environmental degrees of freedom is to be replaced with an ensemble average over noise trajectories. In particular,
Eq.~(\ref{ID}) retains its form.

\section{Open quantum system with Lindblad damping}
Above, we have shown that a classical noise source can be used as an effective description of a decoherence process arising from the microscopic
dynamics of a many-body quantum system within the Born approximation. Below, we consider a physically relevant decoherence source of a
superconducting qubit---a defect coupled to a Markovian environment~\cite{Paladino2002,Falci2002,Falci2005}, and show by tracing out the
environmental degrees of freedom how this model is also mapped analytically to Eq.~\eqref{rtn_master} describing the average system dynamics
under RTN. This central result motivates and further justifies the use of classical noise in master equations of quantum dynamics. The reduction
of the complicated many-body dynamics to the form presented above allows significant conceptual and practical simplification of the problem.

Decoherence mediated by fermionic vacancies in a Markovian environment has been studied in Refs.~\cite{Paladino2002,Falci2002,Falci2005} as a
possible source of RTN and $1/f$ noise in Josephson qubits. In addition, an equivalent defect has been shown to result in the RTN
spectrum~\cite{Shnirman2005}. Motivated by these results, we consider a microscopic quantum system coupled to a defect through the projection
operators on the eigenbasis of the defect. Without loss of generality, the total Hamiltonian of this system and an electron band coupled to the
defect can be expressed as
\begin{align}
H_\textrm{dsb} =&\left(I_\textrm{d}\otimes H_\textrm{s} + \frac{\epsilon}{2}\sigma_{z\textrm{d}}\otimes I_\textrm{s}+\sigma_{z\textrm{d}}\otimes
K_{\textrm{s}}\right)\otimes I_\textrm{b} \nonumber \\ &+H_\textrm{b}+H_\textrm{db},
\end{align}
where the operator~$K_{\textrm{s}}$ of the microscopic quantum system is responsible for the coupling to the defect,~$\sigma_{z\textrm{d}}$ is
the Pauli~$z$ matrix operating in the eigenbasis of the defect,~$\epsilon$ is the energy splitting of the defect, and $H_\textrm{db}$ couples
the defect to an electron band represented by the Hamiltonian~$H_\textrm{b}$. The defect is essentially a bistable quantum fluctuator which can
also be represented by a fermionic vacancy with a population varying between 0 and 1 as electrons tunnel into and out of the band. By
eliminating the band electrons from the dynamics of the defect in the Born-Markov approximation, we obtain effective terms operating on the
defect part of the density operator space represented by the Lindblad operator
\begin{align} \label{damp}
\mathcal{L}_\textrm{d}\{\rho_\textrm{d}\}=\frac{\Gamma_1}{2}\left(2\sigma^-\rho_\textrm{d}\sigma^+
-\sigma^+\sigma^-\rho_\textrm{d}-\rho_\textrm{d}\sigma^+\sigma^-\right)\nonumber\\
+\frac{\Gamma_2}{2}\left(2\sigma^+\rho_\textrm{d}\sigma^- -\sigma^-\sigma^+\rho_\textrm{d}-\rho_\textrm{d}\sigma^-\sigma^+\right).
\end{align}
The first three terms describe the process of electron tunneling out of the defect with rate $\Gamma_1$ and the last three terms describe the
inverse process with rate $\Gamma_2$. The rates obey the detailed balance condition $\Gamma_1/\Gamma_2=\exp[{\epsilon/(k_\textrm{B}T)}]$, where
$T$ is the temperature of the electron band. The equation of motion for the density operator~$\rho_{\textrm{ds}}$ of the microscopic system and
the defect can be written as
\begin{align}\label{lind}
\partial_t\rho_{\textrm{ds}} =&\frac{1}{i\hbar} [I_\textrm{d} \otimes H_\textrm{s}+\frac{\epsilon}{2}\sigma_{z\textrm{d}} \otimes I_\textrm{s}
+\sigma_{z\textrm{d}} \otimes K_{\textrm{s}},\rho_{\textrm{ds}}] \nonumber \\ &+ \left(\mathcal{L}_{\textrm{d}}\otimes
I_\textrm{s}\right)\{\rho_{\textrm{ds}}\}.
\end{align}
By expressing the density operator $\rho_{\textrm{ds}}$ in the form
\begin{align}
\rho_{\textrm{ds}}=
\begin{pmatrix}
  \rhodq_{++} & \rhodq_{+-} \\
  \rhodq_{-+} & \rhodq_{--}
\end{pmatrix},
\end{align}
and inserting it into  Eq.~(\ref{lind}), we obtain
\begin{align}\label{janis}
\partial_t\rho_{\textrm{ds}}=& \frac{1}{i\hbar}\begin{pmatrix}
  [H_\textrm{s},\rhodq_{++}] & [H_\textrm{s},\rhodq_{+-}] \\
\left[H_\textrm{s},\rhodq_{-+}\right]&
[H_\textrm{s},\rhodq_{--}]\end{pmatrix} +\frac{\epsilon}{2i\hbar}
\begin{pmatrix}
  0 & 2\rhodq_{+-} \\
  -2\rhodq_{-+} & 0
\end{pmatrix} \nonumber \\ &
+\frac{1}{i\hbar}
\begin{pmatrix}
 [K_{\textrm{s}},\rhodq_{++}]  & \{K_{\textrm{s}},\rhodq_{+-}\} \\
  -\{K_{\textrm{s}},\rhodq_{-+}\} &-[K_{\textrm{s}},\rhodq_{--}]
\end{pmatrix} \\
&+\frac{\Gamma_1}{2}
\begin{pmatrix}
 -2\rhodq_{++} & -\rhodq_{+-} \\
 -\rhodq_{-+} &  2\rhodq_{++}
\end{pmatrix}
+\frac{\Gamma_2}{2}
\begin{pmatrix}
  2\rhodq_{--} & -\rhodq_{+-} \\
  -\rhodq_{-+} & -2\rhodq_{--}
\end{pmatrix} \nonumber.
\end{align}
Equation~\eqref{janis} shows that the diagonal blocks of~$\rho_{\textrm{ds}}$ decouple from the off-diagonal ones. The diagonal blocks are of
primary importance since the density operator of the system can be expressed as $\rho_\textrm{s} = \trace_\textrm{d} \{\rho_{\textrm{ds}}\} =
\rhodq_{++}+\rhodq_{--}$, and hence its dynamics is completely determined by the two coupled master equations with the same dimension
as~$\rho_\textrm{s}$. In fact, the dynamics of~$\rho_{\textrm{ds}}$ in the high temperature limit corresponds to Eq.~\eqref{rtn_master}, which
is observed by denoting $\rho_{\pm\pm}:=\rho_\pm$, $\Gamma_1=\Gamma_2:=1/\tau_\textrm{c}$, and $H_\pm :=H_\textrm{s}\pm K_{\textrm{s}}$. Thus we
have shown that the decoherence arising from the coupling of the system to the defect can be modeled with classical RTN. For finite
temperatures, Eq.~\eqref{janis} can be recast into the form of Eqs.~\eqref{meq1} showing that the equivalence of classical and quantum dynamics
does not necessarily arise due to the infinite temperature limit.

\section{High-fidelity one-qubit {\footnotesize{NOT}} gates}\label{sec5}
To demonstrate the computational effectiveness of the derived master equations, we implement high-fidelity quantum gates for a qubit under RTN
as in Ref.~\cite{mottonen2006}. We assume that the qubit dynamics can be controlled in the $\sigma_x$ direction and that the noise acts in the
$\sigma_z$ direction with the strength~$\Delta$, \ie, $H_\pm=[a(t)\sigma_x\pm\Delta\sigma_z]/2$ in Eq.~\eqref{rtn_master}. The strength of the
control field~$a(t)$ is assumed to be bounded by~$a_\textrm{max}$.

\begin{figure}[tbh]
\includegraphics[width=200pt]{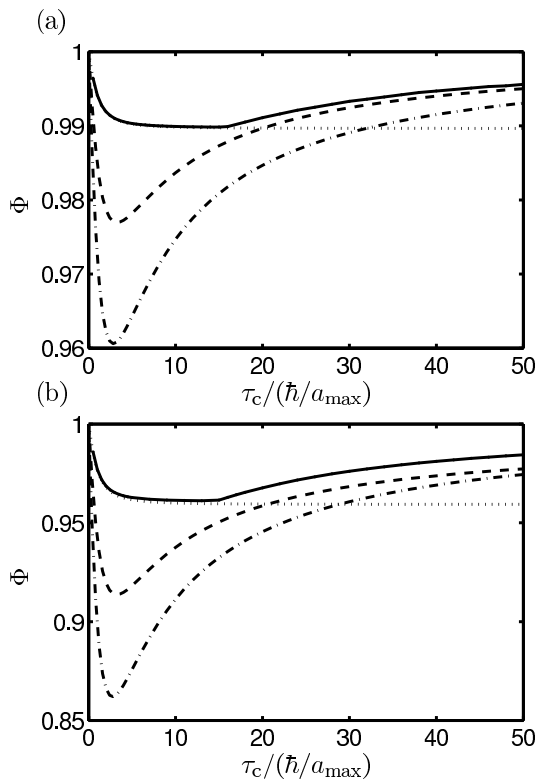} \caption{\label{fig} Fidelities for the {\footnotesize{NOT}}
gate as functions of the correlation time $\tau_{\rm c}$ for a $\pi$-pulse (dotted line), CORPSE (dash-dotted line), short CORPSE (dashed line),
and gradient optimization (solid line). The RTN strength $\Delta$ is set to~(a) $0.125\times a_{\rm max}$ and~(b) $0.25\times a_{\rm max}$. The
gate fidelity for a gate~$U$ is defined as $\langle \textrm{Tr}\{U\rho_0U^\dagger\rho\}\rangle$, where the average is calculated over all pure
initial states~$\rho_0$, see Ref.~\cite{mottonen2006}.}
\end{figure}

\begin{figure}[tbh]
\includegraphics[width=200pt]{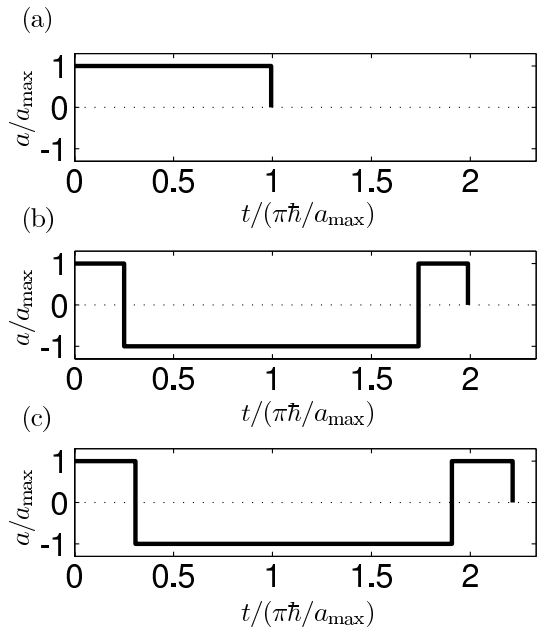} \caption{\label{fig2} Optimized pulse sequences yielding the highest gate fidelities for correlation times (a) $5\hbar/a_\textrm{max}$, (b) $20\hbar/a_\textrm{max}$, and (c) $50\hbar/a_\textrm{max}$. The strength of the noise is chosen to be
$\Delta=0.125a_\textrm{max}$ corresponding to Fig.~\ref{fig}(a).}
\end{figure}

Figure~1 shows gate fidelities~$\Phi$ for the {\footnotesize{NOT}} gate, \ie, the unitary operator~$\sigma_x$, as functions of the noise
correlation time, obtained using composite pulse sequences: $\pi$-pulse, compensation of off-resonance with a pulse sequence (CORPSE), and short
CORPSE~\cite{cummins2000,cummins2003}. To achieve optimized gate fidelity, we employed a variant of the gradient ascent pulse engineering
method~\cite{khaneja2005} of a piecewise constant function~$a(t)$. Figure~2 shows optimized pulse sequences for different correlation times of
the RTN fluctuator. We observe that for short correlation times the optimized sequence is very close to a $\pi$-pulse and for long correlation
times the pulse sequence assumes a shape similar to short CORPSE.

In contrast to similar results presented in Ref.~\cite{mottonen2006} for $\Delta=0.125\times a_\textrm{max}$, the curves in Fig.~\ref{fig} do
not show the statistical errors arising from the finite sampling of the noise. One of the conclusions in Ref.~\cite{mottonen2006} was that the
gradient optimization yields only a marginal improvement to the gate fidelity over the most efficient composite pulse sequence, see
Fig.~\ref{fig}(a). However, we find that this conclusion is valid only for very weak noise, and for example the noise
strength~$\Delta=0.25\times a_\textrm{max}$ employed in Fig.~\ref{fig}(b) suffices to render gradient optimization clearly the most efficient
method considered to fight the decoherence in this system.

\section{Conclusion}
We have shown how classical random telegraph noise can arise directly from coupling of a quantum system to a macroscopic quantum environment or
indirectly through an impurity coupled to a Markovian environment. The average dynamics of the quantum systems were found to coincide under
classical and quantum noise within the Born approximation. Our observations justify the utilization of classical noise as a model for quantum
noise for certain systems. Furthermore, the results presented in this paper introduce a possibility of high-performance pulse optimization for
qubits in noisy environments, and offer an interesting point of view for studies on the effect of noise on quantum systems in general. In the
future, we concentrate on the optimization of multi-qubit operations under noise, study the effects of different quantum baths on the qubit
dynamics, and aim to generalize the obtained equations for continuous and more complicated noise models. The obtained results for the average
quantum dynamics under noise are not restricted to qubits or their multi-level equivalents, qudits, and hence can be employed in studying, \eg,
dilute Bose-Einstein condensates in the noninteracting limit.

\begin{acknowledgments}
We thank the Academy of Finland for financial support. V.\ Bergholm and M.\ M\"ott\"onen acknowledge the Finnish Cultural Foundation, M.\
M\"ott\"onen the Vilho, Yrj\"o, and Kalle V\"ais\"al\"a Foundation, and M.\ M\"ott\"onen and T.\ Ojanen the Magnus Ehrnrooth Foundation for
financial support. S.\ M.\ M. Virtanen, K.\ B.\ Whaley, and J.\ Zhang are appreciated for useful discussions.
\end{acknowledgments}

\bibliographystyle{prsty}
\bibliography{manu}
\end{document}